# The Test of Topological Property of YbB$_6$

**WEIWEI WU**

*JSerra Catholic High School, 26351 Junipero Serra Rd, San Juan Capistrano, CA 92675, USA*

**Abstract:** Topological insulator is a recently discovered class of material with topologically protected surface state. YbB$_6$ is predicted to be moderately correlated Z2 topological insulator similar to SmB6. Here, I experimentally test the resistance property of bulk YbB6 to verify its topological property. By changing the thickness of YbB$_6$, I found out that although the data curves did not completely conform to the theory of topology, the experimental observation to the overall trend showed a similar topological phenomenon.

## 1. Introduction

Topological insulator (TI) [1] is a new class of matter, which has topologically protected surface states that possess unique electronic and spin properties. The relation between the topological order with the electronic correlation already triggered a lot of theoretical considerations and experimental works.

Recently, the Kondo insulator, samarium hexaboride (SmB$_6$), has been predicted to be a Topological Kondo insulator and then studied by many experiments to explore its surface states [2]. The insulator SmB$_6$ is a poor metal at room temperature which transits into an insulator with (limited) conductance at low temperature. When decreasing the temperature, the Kondo lattice metal is transited to a small-gap insulator state, which can protect the metallic surface state. In order to verify its topological property, the separation between bulk and the metallic surface need to be tested. The thickness-dependent measurement is efficient to demonstrate the topologically protected nature of the metallic surface state. With this method, the topological surface state in SmB$_6$ has been proved in 2014 [3].

A related rare-earth hexaboride, YbB$_6$ crystallizes in the CsCl-type structure, like SmB$_6$. It was recently predicted YbB$_6$ is also a moderately correlated topological insulator [4] with a larger band gap than SmB$_6$. However, there are still inconsistencies between theories and experiments in YbB$_6$, thus controversies about whether it's a topological insulator or not [5] continue. Because YbB$_6$ is a promising candidate for TIs, it's of great importance to experimentally test the topological nature of YbB$_6$.

In this paper, I report the thickness-dependent measurement on YbB$_6$ rectangular bulk material. For ideal topological insulators, they deviate from the Ohm's law since they do not have bulk conductance. Therefore, the sample's thickness will not affect its resistance below certain low-temperature points if a three-dimensional (3D) TI transforms from a conventional bulk conductor to an insulator with only surface conduction. I test the resistance of YbB$_6$ bulk with different cross-areas by polishing the crystal. I experimentally found out that below a certain temperature, the resistance of the rectangle sample does not follow the Ohm's law, which may be an indicator of the existence of the topological property.

## 2. Ohm's Law

In Ohm's Law, the resistance equals to the ratio of voltage over current:

$$R = \frac{V}{I} \qquad (1)$$

The continuum for the Ohms law:

$$\Delta V = -\int E \, dl \qquad (2)$$

Where $V$ is the voltage(volt), *i.e.*, the negative integral of the electric field vector(volt/meter) with respect to the path length. Moreover, the current density $J = \sigma E$, where $\sigma$ is the conductivity, and $E$ is the electric field. Therefore, the resistance in terms of the sample's length and cross-area can be derived

$$R = \frac{\rho l}{a} \qquad (3)$$

Where $\rho$ is the resistivity, $l$ is the length, and $a$ is the cross-area. For a traditional metal, the resistance is inverse to the cross area. However, for topological insulator, when cooled down to low temperature, there's a separation between the bulk and surface and only the top layer (two dimensional) of the electrons conduct, which will deviate the three-dimensional Ohm's law. The Ohm's law can be re-investigated on the sample with different thickness at low temperature. And the failure of Ohm's law will prove the sample's deviation from non-topological metal.

## 3. Experimental details

*3.1 Sample preparation*

The experiment begins with the YbB$_6$ sample preparation. First of all, a YbB$_6$ sample is chosen from the sample box, and the hot plate is heated to around 100 °$C$ to melt the crystal wax, which is used to fix the sample on the polisher. Then the sample with the crystal wax is cooled down and solidifies on the polisher, as shown in Fig. 1. The sample is polished into a rectangular using a polish paper. After this, the wax is dissolved with acetone, and the sample, which is ready for measuring is taken out from the acetone.

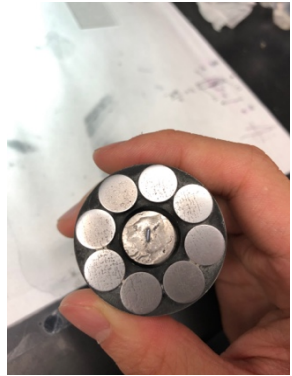

(Fig. 1)

Fig. 1. The YbB$_6$ sample, which is fixed on the polisher by the wax.

*3.2 Measurement procedure*

Firstly, the sample's dimension is measured with a caliper to be [Length: 1321 μm (0.052 in); Width: 901 μm (0.0355 in); Thickness: 457 μm (0.018 in)]. Secondly, four Platinum Tungsten 8% wires are weld onto the sample (Fig 2. a), with the spot welder current output set to 5-10 watt/sec; Finally, Pb37Sn63 is melt to solder the sample onto a puck as in Fig 2. C, using a soldering machine. (the two outer wires connect to Ammeter, and the two inner wires connect to the voltmeter). The puck is put onto the Puck tester, and a multimeter is used to make sure the connection between the sample and the puck is good for double-checking. The puck is then put into a PPMS (Physical Property Measurement System), cooling down to 1.9 K at a rate of 1 K/min, and recording the data as the temperature gradually increase to 60 K at 1 K/min. The sample is taken from PPMS and re-polished to a thickness of 305 μm (0.012 in) after the measurement. Then the sample is soldered onto the puck, and tested with the PPMS again.

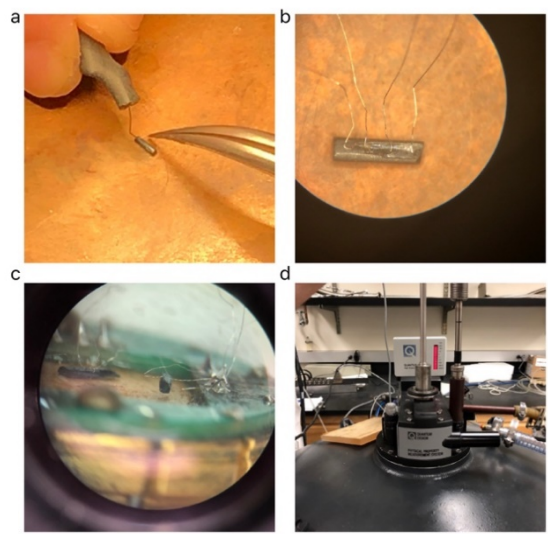

(Fig. 2)

Fig. 2. (a), Welding Platinum Tungsten 8% wires. (b), The YbB$_6$ sample welded with Platinum Tungsten 8% wires. (c), Sample on the puck. (d), PPMS (Physical Property Measurement System)

## 4. Result

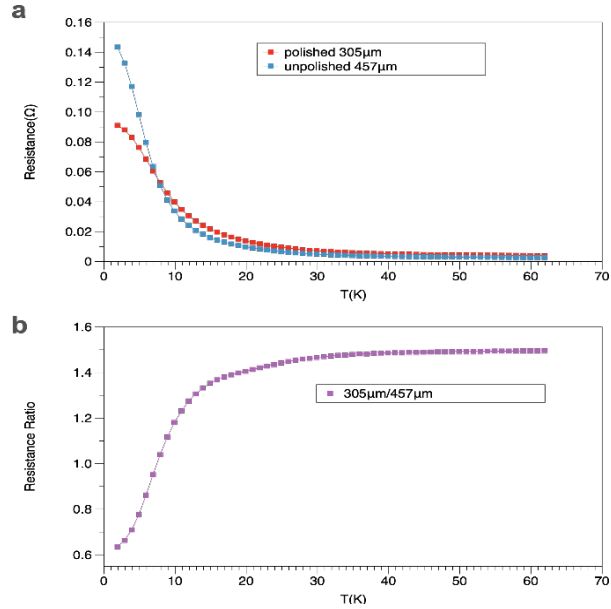

Fig. 3. Experimental results. (a), Resistance versus temperature of YbB6 sample of different thickness. (b), Resistance ratio (polished sample/unpolished sample). As the temperature is below about 10K, the decrease of resistance ratio is an indicator of bulk and surface separation. The ratio below one not agreed with the theory is probably caused by some uncertain issue in the method.

The experimental results are shown in Fig. 3. The resistance of the sample increase as the temperature is cooled down, as shown in Fig. 3(a). This is the property of Kondo insulator. I tested the sample with two different thicknesses, and the ratio of the resistance is calculated and shown in Fig. 3(b). From those results, it is clear that the resistance of the sample is inversely proportional to its cross-area at high-temperature, and the resistance of the sample with both the thickness increases when the temperature decreases.

The resistance of ideal topological materials should disobey the Ohm's law and be irrelevant to their cross-areas since they will become surface conducting. Moreover, because of this surface-conducting property, they should not be influenced by temperature as well. On the contrary, instead of cross-area, the resistance of ordinary materials depends on their cross-areas and obey the Ohm' law. Therefore, on the resistance vs. temperature graph, the topological property should make the curve horizontal because different thicknesses do not affect the material's resistance. Also, on the ratio vs. temperature graph, the curve that has topological phenomenon should tend to 1 at low-temperature points because the resistance values of the sample with surface-conduction will be constant with varying thicknesses. In the experiment, the phenomenon signifies the sample's surface conduction dominates bulk conduction at low temperature, and the cross-area of the sample no longer affects the resistance as Ohm's law claims below that temperature. Based on the observation of the ratio tendency, the results show the sample's topological property although the resistance ratios at low-temperature points trends to 0.6 instead of exactly 1.

## 5.  Conclusion and outlook:

In this experiment, I experimentally tested the YbB6 sample fits the property of Kondo insulator since its resistance increases as the temperature decreases. At low-temperature, although the data curves did not completely conform to the theory of topology, the experimental observation of the overall trend showed a similar topological phenomenon. There are still a few open questions in this experiment, such as the resistance ratios at low-temperature points have large errors from the ideal topological phenomenon. On suspicion, based on the Hysteresis Effect [6], a change in the sample's magnetic moment may be caused in the cooling process and need to be further studied.

### Acknowledgment

I acknowledge the help of Brian Casas in learning the topic and the experiment methods, and
Dr. Bo Li for his assistance in the paper's format and layout.